\documentclass[conference]{IEEEtran}
\IEEEoverridecommandlockouts
\usepackage{cite}
\usepackage{url}
\makeatletter
\newcommand*{\rom}[1]{\expandafter\@slowromancap\romannumeral #1@}
\makeatother
\usepackage{amsmath,amssymb,amsfonts}
\usepackage{algorithmic}
\usepackage{graphicx}
\usepackage{textcomp}
\usepackage{xcolor}
\def\BibTeX{{\rm B\kern-.05em{\sc i\kern-.025em b}\kern-.08em
    T\kern-.1667em\lower.7ex\hbox{E}\kern-.125emX}}
\begin{document}

\title{Voice Activity Detection in presence of background noise using EEG\\
{%\footnotesize \textsuperscript{*}Note: Sub-titles are not captured in Xplore and
}
\thanks{* Equal author contribution.}
}

\author{\IEEEauthorblockN{Gautam Krishna}
\IEEEauthorblockA{\textit{Brain Machine Interface Lab} \\
\textit{The University of Texas at Austin}\\
Austin, Texas \\
}
\and
\IEEEauthorblockN{Co Tran}
\IEEEauthorblockA{\textit{Brain Machine Interface Lab} \\
\textit{The University of Texas at Austin}\\
Austin, Texas \\
}
\and
\IEEEauthorblockN{Mason Carnahan*}
\IEEEauthorblockA{\textit{Brain Machine Interface Lab} \\
\textit{The University of Texas at Austin}\\
Austin, Texas \\
}
\and
\IEEEauthorblockN{Yan Han*}
\IEEEauthorblockA{\textit{Brain Machine Interface Lab} \\
\textit{The University of Texas at Austin}\\
Austin, Texas \\
}
\and
\IEEEauthorblockN{Ahmed H Tewfik}
\IEEEauthorblockA{\textit{Brain Machine Interface Lab} \\
\textit{The University of Texas at Austin}\\
Austin, Texas  \\
}
}

\maketitle

\begin{abstract}
In this paper we demonstrate that performance of voice activity detection (VAD) system operating in presence of background noise can be improved by concatenating acoustic input features with electroencephalography (EEG) features. We also demonstrate that VAD using only EEG features shows better performance than VAD using only acoustic features in presence of background noise.
We implemented a recurrent neural network (RNN) based VAD system and we demonstrate our results for two different data sets recorded in presence of different noise conditions in this paper. 

We finally demonstrate the ability to predict whether a person wish to continue speaking a sentence or not from EEG features.
\end{abstract}

\begin{IEEEkeywords}
electroencephalography (EEG), voice activity detection, deep learning,  technology accessibility 
\end{IEEEkeywords}

\section{Introduction}
Voice activity detection (VAD) system detects presence or absence of human speech. VAD systems are typically used to trigger an automatic speech recognition (ASR) system and helps to improve the performance of the ASR system. Current state of the art VAD systems \cite{moattar2009simple,drugman2015voice} demonstrates good performance in absence of background noise but their performance degrades significantly in presence of high background noise. For example, the experiments carried out in our lab demonstrate that a VAD system fails to detect speech even in presence of a background noise of 40dB. Designing robust VAD systems is crucial to improve the performance of ASR systems operating in presence of high background noise.

Electroencephalography (EEG) is a non invasive way of measuring electrical activity of human brain. EEG sensors are placed on the scalp of a subject to obtain EEG readings. In \cite{krishna2019speech} authors demonstrated isolated speech recognition using EEG features for a limited English vocabulary of four words and five vowels. They also demonstrated that EEG features are less affected by external background noises. In \cite{krishna2019state,krishna20} authors demonstrated continuous speech recognition using EEG features and in \cite{krishna2019state} authors also introduced different types of EEG feature sets.  Motivated by the results demonstrated by authors in \cite{krishna2019speech} we implemented VAD using EEG features to see if the performance of VAD systems operating in presence of high background noise can be improved using EEG features.  

In \cite{kocturova2018eeg} authors proposed methods to perform VAD using EEG signals but they didn't provide any results to support their ideas whereas in this we paper we provide results obtained using EEG signals recorded from real experiments. We demonstrate results using two different data sets in this paper. VAD using EEG features might also help to improve the performance of EEG based ASR systems introduced by authors in \cite{krishna20,krishna2019state}. 

Current VAD systems operate with acoustic input only there by limiting technology accessibility to people with speaking disabilities or people who can't speak at all. VAD using EEG features will allow people with speaking disabilities to use VAD systems, thereby improving technology accessibility. To the best of our knowledge this is the first time a VAD system is demonstrated using only real experimental EEG features.

Voice assistant systems fails to detect break in speech when an user wishes to take a break before continuing the original speech. For example let's imagine an user wishes to utter " what's the weather tomorrow" to the voice assistant system but after he spoke " what's the weather", let's imagine he took a break of few seconds before continuing to utter the next word "tomorrow", then the voice assistant system will treat both the utterances as two separate utterances and will produce wrong response.  In this paper we investigate whether this problem can be corrected using EEG features.

\section{Voice activity detection model}
Our voice activity detection (VAD) model is a recurrent neural network (RNN) based classifier model as shown in Figures 1 and 2. Our model consists of three layers of gated recurrent unit (GRU) \cite{chung2014empirical} with hidden units 128, 32 and 8 respectively when used with first data set and with hidden units 128, 64 and 32 respectively when used the second data set as shown in Figures 1 and 2. Between the GRU layers dropout \cite{srivastava2014dropout} regularization with dropout rate 0.2 is applied. A time distributed dense layer with 4 hidden units is applied after the final GRU layer.  When used with first data set, the time distributed dense layer used sigmoid activation and when used with the second data set, the time distributed dense layer used ReLU \cite{xu2015empirical} activation. The time distributed dense layer is followed by a dense layer of two hidden units which performs an affine transformation. The dense layer output or logits are passed to softmax activation to predict the class probabilities. The number of time steps of the GRU is equal to the product of sampling frequency of the input features and sequence length. In our case there was no fixed value for time steps. 

The GRU based classifier model at every time step predicts whether the given input contains silence or speech. We used one hot vector to label the training data. Target value of 1 was assigned to input feature frame containing speech and a target value of 0 was assigned to the input feature frame containing silence. The models were trained using adam \cite{kingma2014adam} optimizer for 200 epochs to observe loss convergence. The loss function used was categorical cross entropy and batch size was set to one. For each data set we used 80 \% of data to train the model, 10 \% as validation set and remaining 10 \% as test set. The validation set was used to identify the right set of hyper parameters for the model. 

We also tried performing experiments by replacing the GRU layers with temporal convolutional network (TCN) \cite{bai2018empirical} layers but the performance was poor even though TCN training was much faster than training the model with GRU layers. All the scripts were written using Python Keras Deep Learning framework. 

\begin{figure}[h]
\begin{center}
\includegraphics[height=8.5cm, width=\linewidth,trim={0.1cm 0.1cm 0.1cm 0.1cm}]{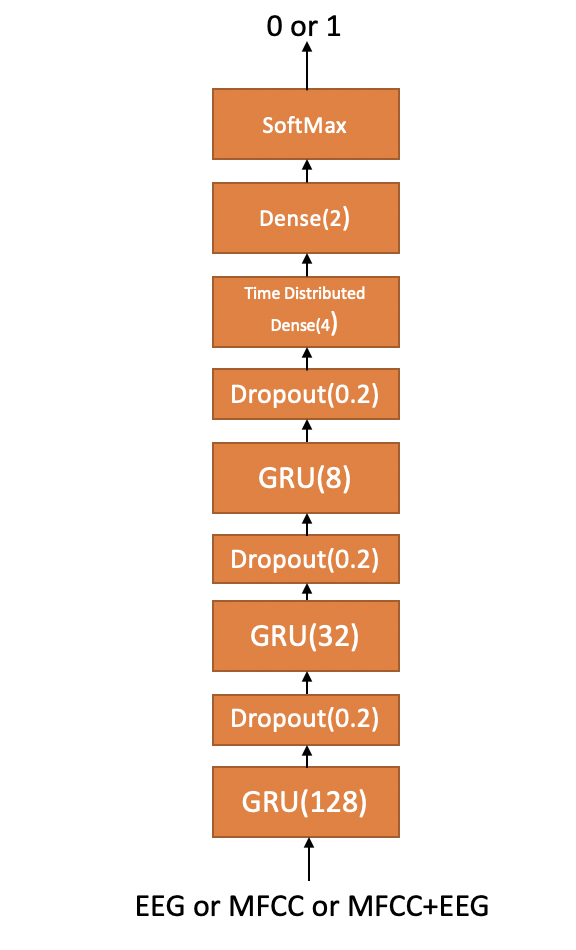}
\caption{VAD Model used with first Data Set} 
\label{1vsall}
\end{center}
\end{figure}

\begin{figure}[h]
\begin{center}
\includegraphics[height=8.5cm, width=\linewidth,trim={0.1cm 0.1cm 0.1cm 0.1cm}]{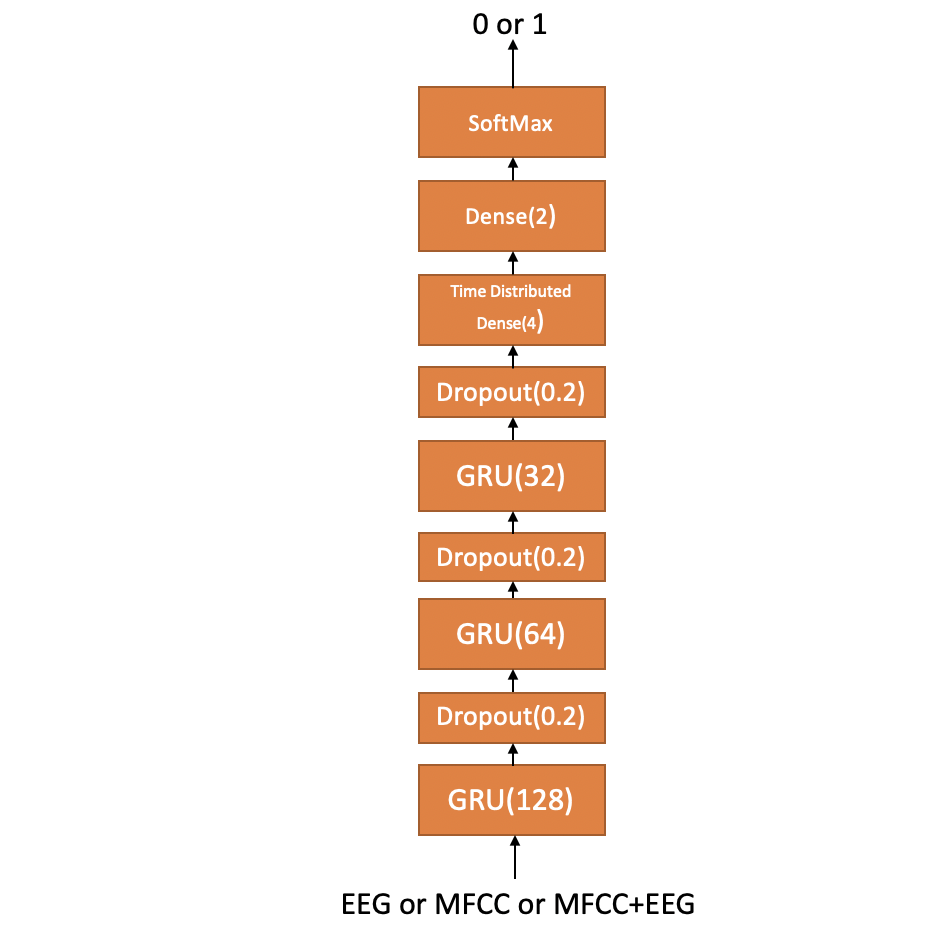}
\caption{VAD Model used with second Data Set} 
\label{1vsall}
\end{center}
\end{figure}

\begin{figure}[h]
\begin{center}
\includegraphics[height=3cm,width=0.25\textwidth,trim={1cm 1cm 1cm 0.1cm},clip]{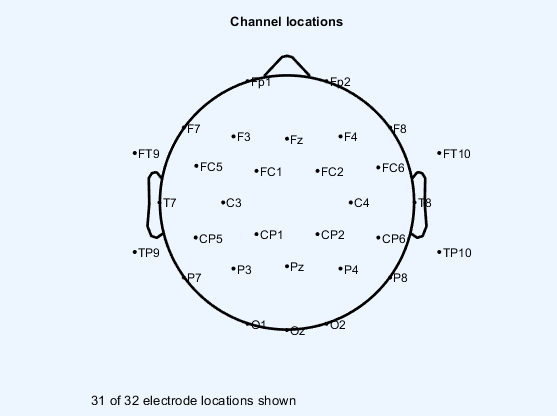}
\caption{EEG channel locations for the cap used in our experiments} 
\label{1vsall}
\end{center}
\end{figure}

\section{Model to predict continue speaking a sentence or not from EEG}

The model consists of two layers of GRU with 64 and 32 hidden units respectively. EEG features are fed to GRU (64) as input. After each GRU layer a dropout regularization with dropout rate 0.2 was applied. The last time step output of GRU(32) is fed into a dense layer with 4 hidden units. The dense layer output is passed to softmax activation to get prediction probabilities.
The model was trained for 200 epochs using adam optimizer. The batch size was set to 100 and categorical cross entropy was used as the loss function for the model. The labels were one hot vector encoded. 80 \% of data was used to train the model, 10 \% as validation set and remaining 10 \% as test set. The model architecture is described in Figure 5. 

\section{Design of Experiments for building the database}

We used two data sets for training and testing VAD model. The first data set was the data set B used by authors in \cite{krishna20} where 8 subjects were asked to speak the first 30 English sentences from USC-TIMIT database \cite{narayanan2014real} and their simultaneous speech and EEG signals were recorded. This data was recorded in presence of background noise of 65dB. Each subject was asked to repeat the experiment two more times. Here the subjects read out loud the English sentences that were shown to them on a computer screen. 

The second data set was the database B used by authors in \cite{krishna2019state} where 15 subjects were asked to listen and speak out the first 9 English sentences from USC-TIMIT database and their simultaneous speech and EEG signals were recorded. This data was recorded in presence of background noise of 50dB. Each subject was asked to repeat the experiment two more times.

For studying the problem of predicting whether a person wish to continue speaking a sentence or not from EEG features, one subject took part in the experiment. The female subject was asked to speak three different related sentences, one unrelated sentence and simultaneous speech and EEG signals were recorded. The four sentences were "what's the weather tomorrow", "what's the weather", "what's the weather today" and "what's the weather macroni". While speaking the sentences "what's the weather today" and "what's the weather tomorrow", the subject took two seconds gap after uttering the phrase "what's the weather" before saying the next word "today" or "tomorrow". 
The subject spoke each sentence 50 times and the data was recorded in absence of external background noise. The subject listened to the utterances first and then speak out the load the utterances.

We used Brain Vision EEG recording hardware. Our EEG cap had 32 wet EEG electrodes including one electrode as ground as shown in Figure 3. We used EEGLab \cite{delorme2004eeglab} to obtain the EEG sensor location mapping. It is based on standard 10-20 EEG sensor placement method for 32 electrodes.

\section{EEG and Speech feature extraction details}

We followed the same EEG and speech preprocessing methods used by authors in \cite{krishna2019speech,krishna20} for all the data sets. 
EEG signals were sampled at 1000Hz and a fourth order IIR band pass filter with cut off frequencies 0.1Hz and 70Hz was applied. A notch filter with cut off frequency 60 Hz was used to remove the power line noise.
EEGlab's \cite{delorme2004eeglab} Independent component analysis (ICA) toolbox was used to remove other biological signal artifacts like electrocardiography (ECG), electromyography (EMG), electrooculography (EOG) etc from the EEG signals. %We used the ICA tool box from Brain Vision for this purpose.  
We extracted five statistical features for EEG, namely root mean square, zero crossing rate,moving window average,kurtosis and power spectral entropy \cite{krishna2019speech,krishna20}. So in total we extracted 31(channels) X 5 or 155 features for EEG signals.The EEG features were extracted at a sampling frequency of 100Hz for each EEG channel.

The recorded speech signal was sampled at 16KHz frequency. We extracted Mel-frequency cepstrum coefficients (MFCC) as features for speech signal.
We extracted MFCC features of dimension 13. The MFCC features were also sampled at 100Hz, same as the sampling frequency of EEG features. 
\begin{figure}[h]
\centering
\includegraphics[height=5cm, width=0.4
\textwidth,trim={0.1cm 0.1cm 0.1cm 0.1cm},clip]{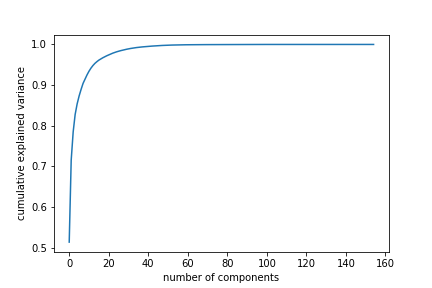}
\caption{Explained variance plot}
\label{1vsall}
\end{figure}

\begin{figure}[h]
\begin{center}
\includegraphics[height=8.5cm, width=\linewidth,trim={0.1cm 0.1cm 0.1cm 0.1cm}]{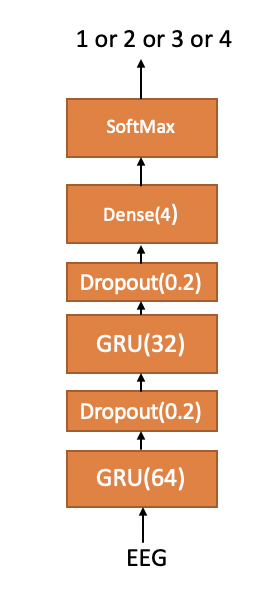}
\caption{ Model used for predicting  continuing speaking a sentence or not from EEG} 
\label{1vsall}
\end{center}
\end{figure}

\section{EEG Feature Dimension Reduction Algorithm Details}
After extracting EEG and acoustic features as explained in the previous section, we used Kernel Principle Component Analysis (KPCA) \cite{mika1999kernel} to denoise the EEG feature space as explained by authors in \cite{krishna20,krishna2019speech}. 
We reduced the 155 EEG features to a dimension of 30 by applying KPCA for all the data sets. We plotted cumulative explained variance versus number of components to identify the right feature dimension as shown in Figure 4. We used KPCA with polynomial kernel of degree 3 \cite{krishna2019speech,krishna20}. 

\section{Results}

We used classification accuracy as performance metric to evaluate the performance of the VAD model on test set data. Classification accuracy can be defined as ratio of number of correct predictions given by the model to total number of predictions given by the model on test set data. 

Table 1 shows the test time results for VAD model when trained and tested using only acoustic or MFCC features, EEG features, concatenation of MFCC and EEG features for both the data sets.
When trained and tested using first data set we observed that concatenating EEG and MFCC features as input gave the highest test time accuracy whereas for the second data set we observed that using only EEG features as input gave the highest test time accuracy. For both the data sets we observed that when the model was trained and tested using only MFCC features as input resulted in lowest test time accuracy. 

Even though data set 1 was recorded in presence of higher background noise than data set 2 we observed that test accuracy using MFCC was higher for data set 1 but data set set 2 MFCC test accuracy was comparable to data set 1. Similar observations were noted for EEG and EEG + MFCC test time accuracy values too.
One possible explanation for these observations might be the nature of the data sets. For data set 1 the subjects were reading out loud the sentences shown on a computer screen where as in data set 2 the subjects first listened to the utterances and then they speak out loud the utterances. The listening utterances might have added additional noise and might have raised the noise level of data set 2 from 50 dB to around 65 dB. 
The EEG recorded in both cases might have slightly different properties and it might depend on the subjects too as each human brain is unique. This needs further exploration and understanding which will be considered for our future work. However for both the data sets we observed that test time accuracy using EEG or EEG + MFCC is higher than using only MFCC as input.

For predicting continue speaking a sentence or not from EEG features, we observed a test accuracy of \textbf{60\%} on test set which was higher than the 35.90\% test accuracy obtained using MFCC features. We used early stopping to prevent over-fitting. 
Test accuracy here is defined as ratio number of correct predictions given by the model to total number of predictions on test set data.

\begin{table}[!ht]
\centering
\begin{tabular}{|l|l|l|l|}
\hline
\textbf{\begin{tabular}[c]{@{}l@{}}Data\\ Set\end{tabular}} & \textbf{\begin{tabular}[c]{@{}l@{}}MFCC\\ (\%accuracy)\end{tabular}} & \textbf{\begin{tabular}[c]{@{}l@{}}EEG\\ (\%accuracy)\end{tabular}} & \multicolumn{1}{c|}{\textbf{\begin{tabular}[c]{@{}c@{}}MFCC\\ +\\ EEG\\ (\%accuracy)\end{tabular}}} \\ \hline
1                                                           & 60.3                                                                 & 83.1                                                                & \textbf{85.7}                                                                                       \\ \hline
2                                                           & 58.07                                                                & \textbf{71.06}                                                      & 64.37                                                                                               \\ \hline
\end{tabular}
\caption{VAD test time results using first and second data set}
\end{table}

\section{Conclusion and Future work}
In this paper we demonstrated voice activity detection (VAD) using only EEG features and we demonstrated that concatenating acoustic features with EEG features as input improves the performance of VAD systems operating in presence of background noise. 

To the best of our knowledge this is the first time a VAD system is demonstrated using only real experimental EEG features.
For future work, we plan to build a much larger speech EEG data base and also perform experiments with data collected from subjects with speaking disabilities.

We will also investigate whether it is possible to improve the test time accuracy of the VAD model and model to predict continue speaking a sentence or not from EEG features by training the models with more number of examples.

\section{Acknowledgement} 
We would like to thank Kerry Loader and Rezwanul Kabir from Dell, Austin, TX for donating us the GPU to train the models used in this work.\\

\bibliographystyle{IEEEtran}

\bibliography{refs}
\end{document}